\documentclass[amsmath,amssymb,aps,nofootinbib,showkeys,superscriptaddress,showpacs,onecolumn,notitlepage]{revtex4-1}
\usepackage{graphicx, epsfig, amssymb} 
\usepackage{amsmath,amssymb,amsfonts,dsfont,mathrsfs,amsthm,mathtools}
\usepackage{bm} 
\usepackage{color}
\usepackage[usenames]{xcolor}
\usepackage{hyperref}
\usepackage{siunitx}
\hypersetup{linktocpage,colorlinks=true,urlcolor=blue,linkcolor=blue,citecolor=blue}  
\usepackage{verbatim}
\usepackage[utf8]{inputenc}
\usepackage{natbib}
\usepackage{float}
\usepackage[font=small,labelfont=bf,singlelinecheck=off,justification=justified]{caption}
\usepackage[font=small,labelfont=bf,singlelinecheck=off,justification=justified]{subcaption}
\usepackage{bigints}

\newcommand{\diff}[1]{\text{d}#1}

\newcommand{\Lag}{\mathscr{L}}

\begin{document}
 
\title{Hairy Taub-NUT/Bolt-AdS solutions in Horndeski theory}

\author{Esteban Arratia}
\email{esarratia@udec.cl}
\affiliation{Departamento de F\'isica, Universidad de Concepci\'on, Casilla, 160-C, Concepci\'on, Chile}

\author{Crist\'obal Corral}
\email{crcorral@unap.cl}
\affiliation{Instituto de Ciencias Exactas y Naturales, Facultad de Ciencias, Universidad Arturo Prat, Avenida Arturo Prat Chac\'on 2120, 1110939, Iquique, Chile}

\author{Jos\'e Figueroa}
\email{josepfigueroa@udec.cl}
\affiliation{Departamento de F\'isica, Universidad de Concepci\'on, Casilla, 160-C, Concepci\'on, Chile}

\author{Leonardo Sanhueza}
\email{lsanhueza@udec.cl}
\affiliation{Departamento de F\'isica, Universidad de Concepci\'on, Casilla, 160-C, Concepci\'on, Chile}

\begin{abstract}
 We present different Taub-NUT/Bolt-anti de Sitter (AdS) solutions in a shift-symmetric sector of Horndeski theory of gravity possessing nonminimal kinetic coupling of scalar fields to the Einstein tensor. In four dimensions, we find locally and asymptotically locally AdS solutions possessing nontrivial scalar field. In higher dimensions, analytical Taub-NUT/Bolt-AdS $p$-branes and solitons are obtained, supported by the existence of $p$ Horndeski scalar fields with axionic profile. The thermodynamic properties are studied through Euclidean methods and it is found that the first law of thermodynamics is satisfied. Moreover, constraints on the parameter space of Horndeski gravity and NUT charge are obtained by demanding positivity of the mass, entropy, and specific heat of the $p$-branes and soliton. We briefly comment on future applications in holography.  
\end{abstract}

\date{\today}

\maketitle

\section{Introduction\label{sec:Introduction}}

Stationary nonsingular Euclidean solutions of the Yang--Mills equations with finite action---also known as instantons---are extremely important in quantum field theories~\cite{Rajaraman:1982is,Shifman:1994ee}. They represent nonperturbative effects at the quantum level and they appear as the leading contribution to physical observables. In the path integral formalism, instantons provide saddle points that can be used to compute transition amplitudes between topologically inequivalent vacua~\cite{Jackiw:1976pf,Callan:1976je}. Their existence in Yang--Mills theories implies a nontrivial structure of the vacuum, which has observable effects at the quantum level~\cite{tHooft:1976rip,tHooft:1976snw}. 

The gravitational analog of instantons has attracted a lot of interest because one would expect that they might play a similar role in quantum gravity (for a review see~\cite{Esposito:1997wt}). A particular class of metrics admitting such interpretation is the Euclidean version of the one found by Taub and Newman--Tamburino--Unti~\cite{Taub:1950ez,Newman:1963yy}---hereafter referred to as Taub-NUT---whose possible observational signatures have been studied in Refs.~\cite{Chakraborty:2017nfu,Chakraborty:2019rna}. Their nontrivial topology is characterized by a nonzero Pontryagin index that gives the number of harmonic self-dual and anti self-dual forms in the middle dimension~\cite{Eguchi:1980jx}. They are connected to the index theorem of the Dirac operator~\cite{Eguchi:1980jx} and they contribute to the axial gravitational anomaly at the quantum level~\cite{AlvarezGaume:1983ig,AlvarezGaume:1984dr}. When the Lorentzian signature and spherical base manifold are considered, the presence of Misner strings (the gravitational analog of the Dirac string) is usually avoided by imposing periodicity on the time coordinate. This condition, however, implies the presence of closed time-like curves and one is usually left with their Euclidean counterpart that is commonly interpreted as a gravitational instanton. It is worth mentioning that the analytical continuation of this spacetime with locally hyperbolic and flat base manifolds admits nonsingular solutions without closed time-like curves~\cite{Ayon-Beato:2015eca,Anabalon:2018rzq,Anabalon:2019equ}. Additionally, the Kruskal extension can be obtained in Lorentzian signature by abandoning the periodicity of the time coordinate~\cite{Miller:1971em}. In fact, recent developments show that the resulting spacetime is geodesically complete, free from pathologies for free-falling observers, and its extended phase thermodynamics is absent of negative thermodynamic volume~\cite{Clement:2015cxa,Clement:2015aka,Kubiznak:2019yiu}; something that is not guaranteed in the Euclidean case~\cite{Johnson:2014pwa,Johnson:2014xza}.

An interesting application of Taub-NUT geometry is the construction of the Kaluza--Klein monopole~\cite{Sorkin:1983ns,Bais:1984xb}, which has motivated different studies from theoretical physics to differential geometry~\cite{Linshaw:2017bpf,Hashemi:2018jbv,Hashemi:2018ujp,Li2019}. The Taub-Bolt solution, on the other hand, is endowed with a horizon and it resembles Euclidean black holes~\cite{Page:1979aj}. Due to this, their thermodynamic properties is an interesting and active research area, ranging from their extended phase structure to holographic heat engines~\cite{Mann:2004mi,Johnson:2014pwa,Johnson:2017ood,Kubiznak:2019yiu,Ballon:2019uha,Durka:2019,Wu:2019pzr}. One of the most remarkable properties of this space is the breakdown of the area/entropy relation by the presence and nontrivial contribution of Misner strings~\cite{Hawking:1998ct,Mann:1999pc,Astefanesei:2004ji}. Additionally, the magnetic charge of dyonic fields on Taub-Bolt spaces is found to be of topological quantum nature while the electric charge is quantized similar as in the case of the Dirac monopole~\cite{Flores-Alfonso:2017kvy}. Holographic aspects of spacetimes possessing NUT charge have been studied in the context of Kerr/CFT correspondence for the Kerr--NUT--AdS black holes~\cite{Sakti:2019udk,Sakti:2019zix,Sakti:2019krw,Sakti:2020jpo}. Indeed, holographic fluids with nontrivial vorticity can be described at the boundary by considering bulk geometries with NUT charge~\cite{Leigh:2011au,Caldarelli:2012cm,Leigh:2012jv,Mukhopadhyay:2013gja}. Some Taub-NUT/Bolt solutions have been found beyond general relativity (GR), for instance, in the low-energy limit of string theory~\cite{Johnson:1994ek,Johnson:1994nj,Burgess:1994kq,Flores-Alfonso:2018adn}, in Lovelock gravity~\cite{Dehghani:2005zm,Dehghani:2006aa,Hendi:2008wq,Flores-Alfonso:2018jra,Corral:2019leh} and beyond~\cite{Bueno:2018uoy}, in conformally coupled theories~\cite{Bardoux:2013swa,Cisterna:2021xxq}, Chern--Simons modified gravity~\cite{Brihaye:2016lsx}, and in gauge theories of gravitation~\cite{Mccrea:1984cr,Bakler:1984cq}.

On the other hand, the low-energy limit of ultraviolet completions of gravity theories~\cite{Polchinski:1998rq,Gambini:2011zz,Berges:2000ew,Loll:2019rdj} usually predicts scalar fields in their particle spectrum~\cite{Svrcek:2006yi,Schwarz:2013uza} as well as higher-curvature corrections to the action principle~\cite{Zwiebach:1985uq}. In fact, the dimensional reduction of the latter also induce scalar fields on the effective theory~\cite{Charmousis:2014mia,Castillo-Felisola:2016kpe}. From the viewpoint of AdS/CFT correspondence, scalar fields are relevant when describing strongly coupled systems~\cite{Maldacena1999,Gubser:1998bc,Witten:1998qj}, since they introduce new sources in the dual quantum field theory at the boundary. They can be used to describe holographic superconductors~\cite{Hartnoll:2008kx,Horowitz:2010gk,Horowitz:2008bn}, as well as mesons through the AdS/QCD program~\cite{Vega:2009zb,Branz:2010ub,Gutsche:2011vb}. These facts show that scalar fields play an important role in our current description of nature.

In gravitation, the most general scalar-tensor theory in four dimensions having similar features of GR, namely diffeomorphism invariance and second-order field equations, is known as Horndeski's theory of gravity~\cite{Horndeski:1974wa}. Its space of solutions is endowed with hairy black holes~\cite{Rinaldi:2012vy,Babichev:2013cya,Anabalon:2013oea,Cisterna:2014nua,Bravo-Gaete:2014haa}, boson stars~\cite{Brihaye:2016lin}, and neutron stars with polytropic and realistic equations of state with and without rotation~\cite{Cisterna:2015yla,Cisterna:2016vdx}. In the context of holography, different aspects of this theory have been explored, for instance, asymptotically Lifshitz black holes with fractional critical exponent~\cite{Bravo-Gaete:2013dca}, holographic entanglement entropy~\cite{Caceres:2017lbr}, and models with spontaneous momentum dissipation~\cite{Jiang:2017imk,Filios:2018xvy,Cisterna:2017jmv,Cisterna:2018hzf,Cisterna:2019uek,Cisterna:2018jqg}, among others~\cite{Liu:2017kml,Li:2018rgn}. Recently, holographic heat engines were analyzed in Ref.~\cite{Hu:2019wwn}, as well as complexity growth~\cite{Feng:2018sqm}, and the holographic AC charge transport~\cite{Wang:2019jyw}. The thermodynamics of these solutions has been explored in Refs.~\cite{Feng:2015wvb,Jiang:2017imk} and it has been shown that Euclidean methods do not yield to the same result for the entropy when compared to the Noether--Wald's formalism~\cite{Feng:2015oea} (see also Ref.~\cite{Hajian:2020dcq}). From a cosmological viewpoint, these theories produce the accelerated expansion of the Universe without the cosmological constant problem (for a review see~\cite{Clifton:2011jh}) and a particular sector is compatible with the constraints imposed by multimessenger astronomy~\cite{Bettoni:2016mij,Ezquiaga:2017ekz,Ezquiaga:2018btd,Baker:2017hug,Sakstein:2017xjx,Heisenberg:2017qka,Kreisch:2017uet,Nojiri:2017hai}. 
 
The aim of this work is to show that Taub-NUT/Bolt-AdS solutions also exist in a particular sector of Horndeski gravity. To this end, we focus on the nonminimal derivative coupling of the scalar field to the Einstein's tensor and found different solutions. First, we obtain the zero mode of the Horndeski scalar propagating in a Taub-NUT background and show that their energy density vanishes at the fixed points, while the norm of the conserved current associated to the shift symmetry diverges. However, it is well-known that this condition can be removed by demanding the radial component of the scalar current to vanish. A finite scalar current on the horizon is a desirable physical condition even in absence of scalar field's backreaction. Then, we consider their backreaction and demand regularity of the conserved current at the fixed points. We obtain a locally Euclidean AdS solution sourced by a self-gravitating scalar field whose on-shell energy-momentum tensor gravitates as a cosmological constant. The thermodynamic properties of the solution are found through Euclidean methods~\cite{Gibbons:1976ue} by introducing proper counterterms that regularize the on-shell action~\cite{Emparan:1999pm}. Afterward, we analyze the behaviour of the metric near the one and two-dimensional set of fixed points and solve the system numerically to find asymptotically locally Euclidean AdS solutions. We compute the mass of the numerical solution analytically by evaluating the renormalized boundary stress-energy tensor at infinity~\cite{Balasubramanian:1999re}. In higher dimensions, we found Taub-NUT/Bolt-AdS analytical $p$-branes by considering a product space between the Hopf-fibered K\"ahler space and $\mathbb{R}^p$, supported by $p$ nonminimally coupled Horndeski scalars with axionic profile. We obtain constraints on the parameter space by demanding positivity on the entropy and specific heat.

The article is organized as follows: In Sec.~\ref{sec:Horndeski}, we describe the particular sector of Horndeski theory we are interested in, together with their field equations. Section~\ref{sec:Taub-NUT} is devoted to give a brief review about the Taub-NUT/Bolt geometry and their regularity conditions. In Sec.~\ref{sec:vacuum}, we revisit the vacuum solution obtained for Horndeski theory with a constant scalar field for the sake of completeness. Then, in Sec.~\ref{sec:zeromode} we solve analytically the zero mode of the Horndeski scalar propagating in a Taub-NUT background. In Sec.~\ref{sec:taubnutscal} we consider their backreaction and present a locally Euclidean AdS space with nontrivial scalar field and an asymptotically locally Euclidean AdS numerical solution, alongside their thermodynamic properties. Next, in Sec.~\ref{sec:branes} we present analytical Taub-NUT/Bolt-AdS $p$-branes and solitons by considering scalar fields with axionic profile. Finally, in Sec.~\ref{sec:conclusions} we present the conclusions and final remarks. 

Throughout this work we use the following conventions: metric signature $(-,+,...,+)$, the Riemann tensor is defined as $R^{\lambda}{}_{\rho\mu\nu} = \partial_\mu\Gamma^{\lambda}{}_{\rho\nu} + ...$, the Ricci tensor is $R_{\mu\nu} = R^{\lambda}{}_{\mu\lambda\nu}$, and the Ricci scalar $R=g^{\mu\nu}R_{\mu\nu}$. Moreover, the shorthand notation $A_{[\mu\nu]}=\tfrac{1}{2}\left(A_{\mu\nu} - A_{\nu\mu} \right)$ and $A_{(\mu\nu)}=\tfrac{1}{2}\left(A_{\mu\nu} + A_{\nu\mu} \right)$ will be used. We work mainly with Euclidean signature, although the action in the next section is given for the Lorentzian case. Thus, one should bear in mind a global minus sign when computing the Euclidean action. 

\section{Horndeski theory of gravity\label{sec:Horndeski}}

Horndeski theory of gravitation is described by the most general action principle constructed out of the metric and a scalar field that leads to second-order field equations~\cite{Horndeski:1974wa}. It has the same symmetries as GR, namely, diffeomorphism and local Lorentz invariance. In this work, we shall focus on a specific shift-symmetric sector---also known as covariant Galileons~\cite{Deffayet:2009wt}---described by the nonminimally coupled action principle~\cite{Liu:2017kml,Li:2018rgn} 
\begin{align}\label{action}
    I[g_{\mu\nu},\phi] &= I_{\rm H} + I_{\rm GHY} + I_{\rm CT},
\end{align}
where
\begin{align}\label{IH}
I_{\rm H}&=  \int_{\mathcal{V}} \diff{^D x} \sqrt{-g}\left[\kappa(R-2 \Lambda)-\frac{1}{2}\left(\alpha g_{\mu \nu}-\beta G_{\mu \nu}\right) \nabla^{\mu} \phi \nabla^{\nu} \phi\right] \equiv \int_{\mathcal{V}} \diff{^D x} \sqrt{-g}\;\Lag_H , \\
\label{IGHY}
I_{\rm GHY} &= 2\kappa\int_{\partial\mathcal{V}}\diff{^{D-1}}\sqrt{-h}\;\mathcal{K},\\
\label{ICT}
I_{\rm CT} &= \int_{\partial\mathcal{V}}\diff{^{D-1}}\sqrt{-h}\bigg[\zeta_1 + \zeta_2\mathcal{R} + \zeta_3 \mathcal{R}_{\mu\nu}\mathcal{R}^{\mu\nu} + \zeta_4 \mathcal{R}^2 + \zeta_5\mathcal{G} +  \zeta_6 h^{\mu\nu}\nabla_\mu\phi\nabla_\nu\phi + \zeta_7 h^{\mu\nu} \nabla_\mu\Box\phi \nabla_\nu\phi +\ldots  \bigg].
\end{align}
Here, $\kappa=(16\pi G)^{-1}$ is the gravitational constant, $G_{\mu\nu} = R_{\mu\nu} - \tfrac{1}{2}g_{\mu\nu} R$ is the Einstein's tensor, while $\alpha$ and $\beta$ are constants that control the canonical and nonminimal kinetic terms of the scalar field, respectively. Moreover, $h_{\mu\nu} = g_{\mu\nu} - n_\mu n_\nu$ is the induced metric on the boundary $\partial \mathcal{V}$, $h$ its determinant, and $n^\mu$ is a space-like unit normal vector with $n_\mu n^\mu = 1$ and $n^\mu h_{\mu\nu}=0$. The extrinsic curvature associated to these hypersurfaces can be expressed as $\mathcal{K}_{\mu\nu} = h^{\lambda}{}_\mu \nabla_\lambda n_\nu$, whose trace is $\mathcal{K}=g^{\mu\nu}\mathcal{K}_{\mu\nu}$. Additionally, $\mathcal{G} = \mathcal{R}^2 - 4\mathcal{R}_{\mu\nu}\mathcal{R}^{\mu\nu} + \mathcal{R}_{\mu\nu\lambda\rho}\mathcal{R}^{\mu\nu\lambda\rho}$ is the Gauss--Bonnet term of the boundary, whose pieces $\mathcal{R}^{\mu}{}_{\nu\lambda\rho}$,  $\mathcal{R}_{\mu\nu}=\mathcal{R}^{\lambda}{}_{\mu\lambda\nu}$, and $\mathcal{R}=g^{\mu\nu}\mathcal{R}_{\mu\nu}$ represent the Riemann tensor, Ricci tensor, and Ricci scalar associated to $h_{\mu\nu}$, respectively, that can be obtained from the Gauss--Codazzi equation and contractions thereof.

The Gibbons--Hawking--York (GHY) term~\cite{York:1972sj,Gibbons:1976ue}, $I_{\rm GHY}$, is added such that the metric sector has a well-posed variational principle with Dirichlet boundary conditions. Indeed, the generalization of this term for scalar-tensor theories of the Horndeski family was found in Ref.~\cite{Padilla:2012ze}. However, when a scalar field with radial dependence is considered, the aforementioned generalization reduces solely to the GHY term for the particular sector of Horndeski theory we are interested in~\cite{Liu:2017kml}. The counterterm, $I_{\rm CT}$, is composed by the Emparan--Johnson--Myers counterterm~\cite{Emparan:1999pm} and additional scalar terms proposed in Ref.~\cite{Li:2018rgn}. These terms are included to regularize the action for configurations with locally AdS asymptotics and they are relevant in the scheme of holographic renormalization. Moreover, it has been recently shown that, in vacuum, it coincides with the methods of Kounterterms when the Weyl tensor vanishes at the boundary~\cite{Anastasiou:2020zwc}.\footnote{In GR, the method of topological renormalization also renders the Euclidean on-shell action and Noether charges finite~\cite{Ciambelli:2020qny}.}

The field equations derived from action~\eqref{action} are obtained by performing stationary variations with respect to the metric and scalar field, giving
\begin{subequations}\label{eom}
\begin{align}\label{eomg}
\mathcal{E}_{\mu\nu} &\equiv G_{\mu\nu} + \Lambda g_{\mu\nu} - \frac{\alpha}{2\kappa}T_{\mu\nu}^{(1)} - \frac{\beta}{2\kappa} T_{\mu\nu}^{(2)} = 0, \\
 \label{eoms}
\mathcal{E} &\equiv \nabla_\mu J^\mu = 0,
\end{align}
\end{subequations}
respectively, where
\begin{align}
\label{tmunu1}
T_{\mu\nu}^{(1)} &= \nabla_\mu\phi\nabla_\nu\phi - \frac{1}{2}g_{\mu\nu}\nabla_\lambda\phi\nabla^\lambda\phi,\\
\notag
T_{\mu\nu}^{(2)} &= \frac{1}{2}\nabla_\mu\phi\nabla_\nu\phi R - 2\nabla_\lambda\phi\nabla_{(\mu}\phi R^{\lambda}{}_{\nu)} - \nabla^\lambda\phi\nabla^\rho\phi R_{\mu\lambda\nu\rho} \\
\notag 
&\quad - \left(\nabla_\mu\nabla_\lambda\phi\right)\left(\nabla_\nu\nabla^\lambda\phi\right)
 + \left(\nabla_\mu\nabla_\nu\phi \right)\Box\phi + \frac{1}{2}G_{\mu\nu}\nabla_\lambda\phi\nabla^\lambda\phi \\
 \label{tmunu2}
&\quad - \frac{1}{2}g_{\mu\nu}\left[\left(\Box\phi\right)^2 - \left(\nabla_\lambda\nabla_\rho\phi\right)\left(\nabla^\lambda\nabla^\rho\phi\right) - 2\nabla^\lambda\phi\nabla^\rho\phi R_{\lambda\rho} \right],\\
\
\label{Jmu}
J^\mu &= \left(\alpha g^{\mu\nu} - \beta G^{\mu\nu}\right)\nabla_\nu\phi.
\end{align}

As mentioned in Sec.~\ref{sec:Introduction}, the space of solutions of this theory has been widely explored in different contexts and a plethora of analytical and numerical results have been found. In this work, we are interested in Euclidean stationary configurations with regular scalar profile at the fixed points. These solutions have not been considered in the literature and they are relevant for their distinct applications exposed in the Introduction. To this end, we focus on Taub-NUT/Bolt-AdS geometry to solve the field equations~\eqref{eom}. The main properties of the latter and principal results are presented in the next section.

\section{Four dimensional Euclidean Taub-NUT/Bolt-AdS solutions\label{sec:Taub-NUT}}

In the following, we focus on the inhomogeneous family of Euclidean metrics proposed in Ref.~\cite{Page:1985bq}.\footnote{Euclidean signature will introduce a global sign flip to the action~\eqref{action}, leaving the field equations unaffected. This change, however, should be taken into account when computing the on-shell Euclidean action for thermodynamic purposes.} They are constructed on complex line bundles over K\"ahler manifolds as
\begin{equation}\label{euclideanmetricansatz}
 \diff{s^2} = f(r)\left(\diff{\tau} + 2n\mathcal{A}_{(k)}\right)^2 + \frac{\diff{r^2}}{h(r)} + (r^2 - n^2)\diff{\Sigma^2_{(k)}}, 
\end{equation}
where $\tau$ is the Euclidean time, $n$ is the NUT charge, and $\diff{\Sigma^2_{(k)}}$ denotes the line element of the base manifold with constant curvature $k=\pm1,0$, for spherical, hyperbolic, and flat sections, respectively. These bundles are labeled by the first Chern number at infinity which is related to the NUT charge~\cite{Hawking:1998ct}. The K\"ahler potential $\mathcal{A}_{(k)}$ for the different base manifolds can be expressed as (see for example Ref.~\cite{Bardoux:2013swa})
    \begin{align}\label{kahlerpotential}
 \mathcal{A}_{(k)} &= \left\{
    \begin{aligned}
      \cos\theta\diff{\varphi}&  &\mbox{when} & &\diff{\Sigma^2_{(k=1)}} &= \diff{\theta^2} + \sin^2\theta\diff{\varphi^2},\\
      \tfrac{1}{2}\left(\theta\diff{\varphi} - \varphi\diff{\theta}\right) &  &\mbox{when} & &\diff{\Sigma^2_{(k=0)}} &= \diff{\theta^2} + \diff{\varphi^2},\\
       \cosh\theta\diff{\varphi}& &\mbox{when} & & \diff{\Sigma^2_{(k=-1)}} &= \diff{\theta^2} + \sinh^2\theta\diff{\varphi^2}.
    \end{aligned}
  \right.
\end{align}
The real fundamental form $\Omega_{(k)}$ associated to the K\"ahler metric of the base manifold is defined through $\Omega_{(k)}=\diff{\mathcal{A}_{(k)}}$. Even though the line element $\diff{\Sigma}_{(k=1)}^2$ is locally isomorphic to the one of $\mathbb{CP}^1$, the equivalence between $\mathbb{CP}^p$ and $\mathbb{S}^{2p}$ does not hold for $p \geq 2$, since no hyperspheres admit Kahler structures in that case~\cite{Bishop:1965}. The magnetic part of the Weyl tensor is turned on by NUT charge and the latter is generically interpreted as the magnetic mass of the geometry~\cite{LyndenBell:1996xj,Bicak:2000ea,Araneda:2016iiy,Flores-Alfonso:2018jra}. 

In order to have regular Euclidean hypersurfaces with either zero or two-dimensional fixed point---also referred to as nuts or bolts~\cite{Page:1979aj,Bais:1984xb}, respectively---, the following conditions on the metric functions must met
\begin{subequations}\label{boundaryconditions}
\begin{align}\label{fornut}
    \mbox{For NUT:} \;\;\;\;\; f(r)\big|_{r=n} &= 0 \;\;\; \mbox{and}\;\;\; \sqrt{f'(r)h'(r)}\Big|_{r=n} \;\,= \frac{4\pi}{\beta_\tau}, \\
    \label{forbolt}
    \mbox{For Bolt:} \;\;\;\; f(r)\big|_{r=r_b} &= 0 \;\;\; \mbox{and}\;\;\; \sqrt{f'(r)h'(r)}\Big|_{r=r_b} = \frac{4\pi}{\beta_\tau}, 
\end{align}
\end{subequations}
as well as the vanishing of the metric function $h(r)$ at the fixed points. Here, prime denotes differentiation with respect to $r$, $\beta_\tau$ represents the period of the Euclidean time, and the horizon of the Taub-Bolt solution satisfies $r_b>n$. On the other hand, when $k=1$, unobservability of Misner strings imposes that $\beta_\tau = 8\pi n$. This condition, when combined with the absence of conical singularity ensured by the conditions~\eqref{boundaryconditions}, relates $r_b$ with the NUT charge. For the sake of completeness, we review the vacuum case of Horndeski theory~\eqref{action} in the next section.

\subsection{Vacuum case\label{sec:vacuum}}

It is instructive to analyze first the vacuum solution, namely $\phi(r)=$ constant, such that the theory reduces to GR. In this case, the Klein--Gordon equation is automatically satisfied and the metric functions that solve the Einstein's field equations are
\begin{align}\label{fsolGR}
    f(r) &= h(r) =  k\left(\frac{r^2 + n^2}{r^2-n^2}\right) -\frac{2MG r}{r^2-n^2} - \frac{\Lambda}{3}\frac{\left(r^4 - 6n^2 r^2 - 3n^4 \right)}{r^2-n^2} .
\end{align}
Here, $M$ is an integration constant related to the mass, as it can be obtained from the Euclidean on-shell action~\cite{Emparan:1999pm} or by evaluating the regularized boundary stress-energy tensor at infinity~\cite{Balasubramanian:1999re}. This metric reduces to the Schwarzschild-AdS black hole with Euclidean signature in the limit $n\to0$. 

For the Taub-NUT case, the regularity condition~\eqref{fornut} implies that 
\begin{align}\label{MnutGR}
    MG &= kn + \frac{4}{3}\Lambda n^3,
\end{align}
and it is straightforward to read $\beta_\tau = 8\pi n/k$. The case $k=0$ leads to noncompactness of the Euclidean time, while the case $k=-1$ has a horizon before $r=n$, in contradiction with the NUT hypothesis. For $k=1$, the space is topologically equivalent to $\mathbb{R}^4$ since it represents the Hopf fibration of the Euclidean time over $\mathbb{S}^2$. In this case, the Weyl tensor is globally self-dual and the solution has zero mass when the Pontryagin density is added (see Ref.~\cite{Araneda:2016iiy}). Thus, it can be regarded as a ground state labeled by the NUT charge, similar to self-dual instantons in Yang--Mills theory~\cite{Belavin:1975fg}.

In the Taub-Bolt case, on the other hand, the solution is endowed with a horizon at $r=r_b>n$~\cite{Page:1979aj} and the regularity conditions~\eqref{forbolt} imply that the integration constant is fixed as
\begin{align}\label{MboltGR}
    MG &= \frac{k}{2r_b}\left(r_b^2 + n^2 \right) - \frac{\Lambda\left(r_b^4-6n^2r_b^2 - 3n^4 \right)}{6r_b}.
\end{align}
Here, however, the Weyl tensor is no longer globally self-dual even without the cosmological constant. It is direct to check that the period of the Euclidean time in this case is given by 
\begin{align}
    \beta_\tau &= \frac{4\pi r_b}{k-\Lambda\left(r_b^2-n^2 \right)}.
\end{align}
Additionally, when $k=1$, the unobservability of Misner strings demands $\beta_\tau = 8\pi n$ which, in turn, impose a relation between $r_b$ and $n$, that is
\begin{align}\label{rbsol}
   r_b &= - \frac{1\pm\sqrt{1+16\Lambda n^2\left(\Lambda n^2 + 1 \right)}}{4\Lambda n}.
\end{align}
The condition $r_b>n$ must be met alongside the positivity of the argument of the square root of Eq.~\eqref{rbsol}. These two restrictions impose a range of values for the NUT charge such that the Taub-Bolt solution exists. 

In the next section, we obtain analytically the zero mode of the Horndeski scalar propagating on the Taub-NUT/Bolt background by solving the Klein--Gordon equation for a nonminimally coupled scalar field to the Einstein's tensor.

\subsection{Horndeski scalar zero mode in the Taub-NUT background\label{sec:zeromode}}

In the test field limit, the scalar field's zero mode with radial profile, i.e. $\phi=\phi(r)$, satisfies a Klein--Gordon equation~\eqref{eoms} on the Taub-NUT/Bolt-AdS background that is 
\begin{align}
     \frac{\alpha+\Lambda\beta}{r^2-n^2}\frac{\diff{}}{\diff{r}}\left[\left(r^2-n^2 \right)f\phi' \right] = 0,
\end{align}
where $f=f(r)$ is given in Eq.~\eqref{fsolGR}. This equation admits the first integral of motion
\begin{align}
    \phi'(r) &= \frac{\mu}{\left(r^2-n^2 \right)f},
\end{align}
where $\mu$ is an integration constant. 

The energy-momentum tensor of the Horndeski scalar can be defined as $T_{\mu\nu} = \alpha T_{\mu\nu}^{(1)} + \beta T_{\mu\nu}^{(2)}$. Projecting onto the temporal component of the latter through the orthonormal frame $e^{\mu}{}_a$, i.e. $T_{ab} = e^{\mu}{}_a e^{\nu}{}_b T_{\mu\nu}$, the energy density of the zero mode on the Taub-NUT-AdS background is given by
\begin{align}
\begin{aligned}
    \rho(r) = T_{00} = f^{-1}T_{tt} =& \frac{\mu^2}{18f\left(r^2-n^2 \right)^5 }\Bigg\{2\left[3r^2+2n^2 \right]\left[k\left(r^2+n^2 \right)-2GMr\right] -\left(r^2-n^2 \right)^3 \\ 
    &\quad - \frac{\Lambda}{3}\left(9r^6-41n^2r^4-33n^4r^2-15n^6 \right) \Bigg\}. 
\end{aligned}
\end{align}
It is straightforward to check that it vanishes at the zero or two-dimensional fixed points, i.e. $r\to n$ or $r\to r_b$, for nuts and bolts, respectively. Asymptotically, their energy-momentum tensor behaves as the cosmological constant since $\rho(r)\sim - \Lambda r^2/3$. However, the norm of the scalar current given by Eq.~\eqref{eoms} is 
divergent at the fixed points. It is well-known that this problem can be circumvented by demanding the vanishing of the radial component of the current for a self-gravitating scalar field (see for instance Ref.~\cite{Rinaldi:2012vy,Babichev:2013cya,Anabalon:2013oea,Charmousis:2015aya}).

In the following, we consider the backreaction of Horndeski scalars to obtain analytical and numerical solutions to the field equations~\eqref{eom} by imposing regularity of the scalar current at the fixed points.

\subsection{Taub-NUT/Bolt solutions with self-gravitating scalar field\label{sec:taubnutscal}}

In order to find Euclidean solutions with self-gravitating scalar field, we focus on the particular case $k=1$ and consider a radial ansatz $\phi=\phi(r)$. To ensure regularity of scalar current's norm at the fixed points, the restriction $J^r = 0$ must be met~\cite{Charmousis:2015aya}, that is,
\begin{align}\label{scalarcond}
    J^r &= \left(\alpha g^{rr} - \beta G^{rr} \right)\phi' = 0.
\end{align}
Thus, nontriviality of the scalar field demands a relation between the metric functions that can be read from the branch $\left(\alpha g^{rr} - \beta G^{rr}\right) = 0$. This allows one to evade the no-hair theorem of Ref.~\cite{Hui:2012qt} and it relates the metric functions according to
\begin{align}\label{hsolnut}
    h(r) &=  -\frac{f\left\{f\beta n^2 - \left(r^2-n^2 \right)\left[\beta + \alpha\left(r^2-n^2 \right)\right] \right\}}{r\beta\left[\left(r^2 -n^2\right)f' + rf \right]}.
\end{align}
This condition automatically solves the Klein--Gordon equation~\eqref{eoms}. Then, $\mathcal{E}_{rr}=0$ is solved by the scalar field 
\begin{align}\label{phisolnut}
\phi'\,^2(r) &= -\frac{2r\kappa\left(r^2-n^2\right)^2\left(\Lambda\beta+\alpha \right)\left[\left(r^2 -n^2\right)f' + fr \right]}{f\left[\alpha\left(r^2 -n^2\right)^2 + \beta\left(r^2-n^2 \right) - \beta fn^2 \right]^2}.
\end{align}
Replacing~\eqref{hsolnut} and~\eqref{phisolnut} into $\mathcal{E}_{tt}=0$, we obtain the master equation for the remaining metric function $f(r)$ as
\begin{align}\notag
 0&=\Big[f\beta n^2-\left(r^2-n^2\right)\left(\beta+\alpha\left[r^2-n^2\right]\right)\Big]\Big[\left(\Lambda\beta-\alpha\right)\left(r^2-n^2\right)^2-2\beta\left\{r^2-n^2\left(f+1\right)\right\}\Big]f''\\
 \notag
 & -\beta n^2 \Big[\left(\Lambda\beta+3\alpha\right)\left(r^2-n^2\right)^2+2\beta\left\{r^2-n^2\left(f+1\right)\right\}\Big]f'^2 + \frac{1}{r\left(r^2-n^2 \right)}\bigg[ 2\beta^2 n^4 \left(n^2+7r^2\right)f^2\\
 \notag
 &-\beta n^2\left(r^2-n^2\right)\Big\{\left(\Lambda\beta-3\alpha\right)n^4+\left(4\beta-8\Lambda\beta r^2-18\alpha r^2\right)n^2+\left(7\Lambda r^4+18 r^2\right)\beta+21\alpha r^4\Big\}f\\
 \notag
 &+\left(r^2-n^2\right)\Big\{\Lambda\beta^2\left(2r^6-n^6+4n^4r^2-5n^2r^4\right)-\alpha n^2\left(\Lambda\beta-\alpha\right)\left(r^2-n^2\right)^3\\
 \notag
 &+\beta\left[3\alpha\left(2r^6+n^6-3n^2r^4\right)+2\beta\left(2r^4-n^4-n^2r^2\right)\right]\Big\}\bigg]f' + \frac{1}{r\left(r^2-n^2\right)^2}\bigg[ 8f^3\beta^2 n^4 r^3 \\
 \notag 
 &  -4n^2\beta r^3\left(r^2-n^2\right)\left[2\Lambda\beta\left(r^2-n^2\right)+3\alpha\left(r^2-n^2\right)+2\beta\right]f^2 + 2r^3\left(r^2-n^2 \right)^3\Big\{ 2\Lambda\beta^2\\
 \label{mastereq}
 & +\alpha\left(\Lambda\beta-\alpha \right)\left(r^2-n^2 \right)\Big\}f \bigg].
\end{align}
This is a second-order nonlinear equation that is rather challenging to solve analytically. In particular, when $n\to0$, it is solved by the Euclidean version of the black hole reported in Refs.~\cite{Anabalon:2013oea,Babichev:2013cya}. When the NUT charge is nonvanishing, further progress can be done numerically. However, before presenting the numerical solutions, we show the existence of a nontrivial scalar profile that generates a locally Euclidean AdS ground state with an effective curvature radius. This configuration is discussed next.

\subsubsection{Locally Euclidean Taub-NUT-AdS solution with nontrivial scalar field}

Although Eq.~\eqref{mastereq} looks challenging, it admits the following analytical solution with a nontrivial scalar profile
\begin{subequations}\label{maxsym}
\begin{align}\label{fmaxsym}
    f(r) &= \frac{r^2 - n^2}{4n^2}, \\
    \label{hmaxsym}
    h(r) &= \frac{\left[r^2-n^2\right]\left[3\beta+4\alpha\left(r^2-n^2 \right) \right]}{12\beta r^2}, \\
    \label{scalarsolads}
    \phi(r) &= \phi_0 \pm \frac{\sqrt{-6\kappa\left(\Lambda\beta + \alpha \right)}}{2\alpha}  \ln\left[3\beta + 4\alpha\left(r^2-n^2 \right) \right],
\end{align}
\end{subequations}
where $\phi_0$ is an integration constant that, by virtue of the shift symmetry $\phi\to\phi+\mbox{constant}$, can be set to zero without loss of generality. Reality conditions on the scalar field implies that $\Lambda\beta + \alpha < 0$. Moreover, it is direct to see that this solution has a nut since it vanishes at $r=n$ and the period of the Euclidean time is given by $\beta_\tau=8\pi n$. Thus, their Hawking's temperature is $T= \left(8\pi n\right)^{-1}$. This metric has ten Killing vectors and it is locally AdS$_4$ with Euclidean signature, whose Riemann tensor is 
\begin{align}\label{Riemannmaxsym}
    R^{\lambda\rho}{}_{\mu\nu} &= -\frac{2\alpha}{3\beta}\delta^{\lambda}_{[\mu}\delta^{\rho}_{\nu]}.
\end{align}
The effective curvature radius can be read off from the last expression as $\ell_{\rm eff}^{-2} = \tfrac{\alpha}{3\beta}$ and it is direct to check that it has a vanishing Pontryagin density. On the other hand, the stress-energy tensor of the Horndeski scalar gravitates on-shell as a cosmological constant, i.e.
\begin{align}
    \left(T_{\mu\nu}^{(1)} + T_{\mu\nu}^{(2)} \right)\Big|_{\rm on-shell} &= \left(\frac{\alpha}{\beta} + \Lambda \right)g_{\mu\nu}.
\end{align}
This behaviour renders the one of Ref.~\cite{Bravo-Gaete:2014haa} and it generalizes their scalar field's configuration in presence of the NUT charge. 

The thermodynamics of this solution can be obtained from the quantum statistical relation $\ln Z \approx - I$ to lowest order in the saddle-point approximation, where $Z$ is the partition function and $I$ is the Euclidean on-shell action~\cite{Gibbons:1976ue}. The relevant quantity for computing the latter is the Horndeski Lagrangian evaluated at the Taub-NUT ansatz, i.e.  
\begin{align}\notag
    \Lag_H &= \kappa\bigg[-\frac{f''\,h}{f} + \frac{f'^2 h}{2f^2} - \left(\frac{h'}{2f} + \frac{2hr}{f\left(r^2-n^2 \right)}\right)f' - \frac{2h'r}{r^2-n^2}  + \frac{2\left[\left(r^2 - n^2\right)\left(1-h \right) - n^2\left(f-h\right)  \right]}{\left(r^2-n^2 \right)^2} -2\Lambda \bigg]\\
    \label{LagH}
    &\quad - \frac{\alpha}{2}h\phi'^2  + \frac{\beta h\phi'^2}{2f\left(r^2-n^2 \right)}\left[f'hr - f + \frac{fhr^2 + f^2n^2}{r^2-n^2} \right].
\end{align}
To calculate the GHY term and the counterterms that regularize solutions with local AdS asymptotics, the extrinsic and intrinsic curvature scalars associated to a space-like unit vector $n^\mu$ are needed, which are respectively given by
\begin{align}
     \mathcal{K} = \frac{\sqrt{h}\left[\left(r^2-n^2 \right)f' + 4fr \right]}{2f\left(r^2-n^2 \right)} \;\;\;\;\; \mbox{and} \;\;\;\;\; \mathcal{R} = \frac{2\left[r^2 - n^2\left(1+f \right) \right]}{\left(r^2-n^2 \right)^2}.
\end{align}
The higher-curvature contributions and scalar derivative terms are subleading in $D=4$ for the asymptotic expansion and they do not contribute in this case. Moreover, the boundary Gauss--Bonnet term vanishes identically for $D=4$ and it contributes only if $D\geq5$. The Euclidean on-shell action evaluated at the solution~\eqref{maxsym} is finite provided that the counterterms in Eq.~\eqref{ICT} are chosen such that
\begin{align}\label{ctmaxsym}
    \zeta_1 &= \frac{2\kappa\left(\Lambda\beta-\alpha\right)}{\sqrt{3\alpha\beta}} \;\;\;\;\; \mbox{and} \;\;\;\;\; \zeta_2 = - \frac{\kappa\sqrt{3\beta}\left(\Lambda\beta+3\alpha\right)}{2\alpha^{3/2}}.
\end{align}
These values of the parameters guarantee that the action is finite as $r\to\infty$. Thus, the regularized Euclidean action for the solution~\eqref{maxsym} is
\begin{align}\label{Imaxsym}
    I_{\rm reg} = \frac{3\pi\beta\left(\Lambda\beta + 2\alpha \right)}{2G\alpha^2}.
\end{align}
The free energy, mass, entropy, and specific heat can be obtained from this expression as
\begin{align}
\mathcal{F} = \beta_\tau^{-1} I_{\rm reg}, \;\;\;\;\; \mathcal{M} = \frac{\partial I_{\rm reg}}{\partial \beta_\tau}, \;\;\;\;\; \mathcal{S} = \beta_\tau\frac{\partial I_{\rm reg}}{\partial\beta_\tau} - I_{\rm reg}, \;\;\;\;\; \mathcal{C} = - \beta_\tau\frac{\partial S}{\partial \beta_\tau},    
\end{align}
respectively. Notice that the Euclidean action~\eqref{Imaxsym} does not depend on the period of the Euclidean time. Therefore, its thermodynamic mass $\mathcal{M}$ vanishes. Even more, since the solution is locally AdS, its Weyl tensor vanishes identically. This implies that its mass is zero, similar to what happens with the Taub-NUT solution in Einstein gravity when the Gauss--Bonnet and Pontryagin densities are included~\cite{Araneda:2016iiy}. We interpret this configuration as a ground state which is disconnected of global AdS by the presence of a nontrivial stealth-like scalar field. Thus, if a background subtraction prescription would have been used, this solution may regularize the Euclidean action as in Ref.~\cite{Hawking:1998ct}.

On the other hand, it is well-known that the Misner string contributes nontrivially to the entropy in spacetimes with spherical section, as in the case under study here (see for instance Refs.~\cite{Mann:1999pc,Ciambelli:2020qny}). For the solution in Eq.~\eqref{maxsym}, the entropy arises purely from the Misner string, giving
\begin{align}\label{entropymax}
    \mathcal{S} &= -\frac{3\pi\beta\left(\Lambda\beta + 2\alpha \right)}{2G\alpha^2}.
\end{align}
One can notice that the entropy does not depend on the period of the Euclidean time either. This implies that the first law of thermodynamics is trivially satisfied and the specific heat of this solution vanishes. The free energy is proportional to the Hawking's temperature as it can be seen from $\mathcal{F} = \beta_\tau^{-1} I_{\rm reg}$. Thus, in order for the entropy to be positive definite, the constraint $\Lambda\beta+2\alpha<0$ must be met. Compatibility of the latter with real scalar fields and the absence of ghosts implies that $\Lambda<0$, $\beta>0$, and $0<\alpha< - \Lambda\beta/2$. 

In the next section, we obtain Euclidean numerical asymptotically locally AdS solutions with nontrivial scalar profiles in Horndeski gravity. To this end, first we study first series solution near the fixed points of nuts and bolts, respectively.


\subsubsection{Asymptotically locally Euclidean AdS Taub-NUT/Bolt solutions}

In order to solve the system numerically, we need to obtain initial values compatible with the field equations. For NUT, the regularity conditions~\eqref{fornut} imply that the metric function admits a series expansion near $r=n$ as
\begin{align}
    f(r) = f_1\left[\left(r-n\right) + f_2\left(r-n\right)^2 + f_3\left(r-n\right)^3 + f_4\left(r-n\right)^4 +  \ldots \right].
\end{align}
Inserting this expression into Eq.~\eqref{mastereq} an solving for each order of $(r-n)$, it is found that $f_1$ is completely determined by the field equations, giving $f_1 = \tfrac{1}{2n}$. Moreover, every coefficient $f_i$ with $i\geq3$ can be obtained recursively in terms of $f_2$ that remains as a free parameter. In GR, the latter is related to the mass and, once the condition~\eqref{fornut} is imposed, it is fixed in terms of the NUT charge. In this case, $f_2$ remains arbitrary even after the NUT condition. Therefore, we conclude that this solution is endowed with an additional free parameter that is interpreted as a scalar hair. On the other hand, the values of the first coefficients are explicitly given by
\begin{align}
    f_3 &= \frac{1}{48n^2} \left[40\Lambda n^2 + 30n f_2 + 9 \right]\left[2n f_2 - 1 \right],\\
    \notag
    f_4 &= \frac{1}{144\beta^2n^3}\Big[160\Lambda^2\beta^2n^4-128\Lambda\alpha\beta n^4+336\Lambda\beta^2n^3f_2-64\alpha^2n^4+126\beta^2n^2f_2^2\\
    &\quad +12\Lambda\beta^2n^2+9\beta^2nf_2-18\beta^2\Big]\left[2nf_2-1\right],
\end{align}
an so on. If $f_2 = 1/(2n)$, all the coefficients $f_i$ vanishes for $i\geq3$, and this solution reduces to the ground state discussed in the previous section [cf. Eq.~\eqref{maxsym}]. Thus, in order for the numerical solutions to deviate from Eq.~\eqref{maxsym} while approaching to~\eqref{Riemannmaxsym} asymptotically, $f_2$ must be subjected to the condition $f_2 \neq 1/(2n)$. Additionally, the period of the Euclidean time in this case is given by $\beta_\tau = 8\pi n$.

\begin{figure}[t]
\begin{subfigure}[b]{0.5\textwidth}
    \centering
    \includegraphics[scale=0.45]{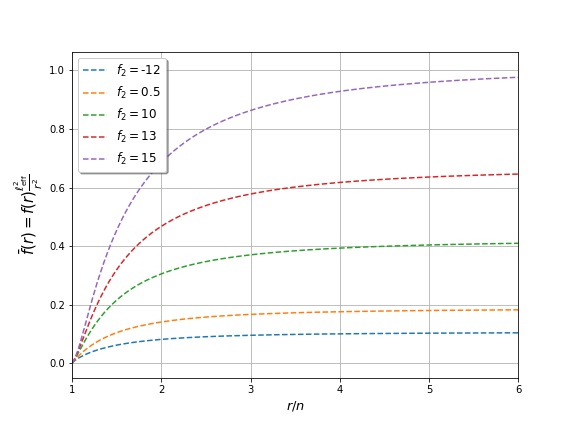}
\end{subfigure}~
\begin{subfigure}[b]{0.5\textwidth}
    \centering
    \includegraphics[scale=0.45]{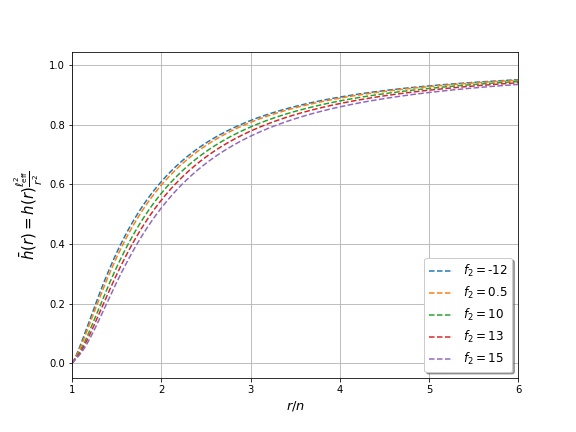}
\end{subfigure}
\caption{Numerical solution of Eq.~\eqref{mastereq} with $\alpha=4$, $\beta=1$, $\Lambda=-10$, $n=1$, and initial conditions~\eqref{icnut}. Here, we defined $\bar{f}(r)=f(r)\tfrac{\ell_{\rm eff}^2}{r^2}$ and $\bar{h}(r)=h(r)\tfrac{\ell^2_{\rm eff}}{r^2}$. The function $\bar{f}(r)$ reaches different asymptotic values when varying $f_2$, which is related to the fact that $\ell_{\infty}$ remains arbitrary in Eq.~\eqref{fasymnut}. On the other hand, asymptotically $h(r)= r^2/\ell_{\rm eff}^2+\ldots$ for arbitrary values of $\ell_\infty$ and $\mu_1$, with $\ell_{\rm eff}^{-2}=\tfrac{\alpha}{3\beta}$. Thus, $\bar{h}(r)\to1$ as $r\to\infty$. The case $f_2=0.5$ represents the ground state given by Eq.~\eqref{maxsym}.}
\label{barfnut}
\end{figure}

\begin{figure}[t]
\begin{subfigure}[b]{0.5\textwidth}
    \centering
    \includegraphics[scale=0.45]{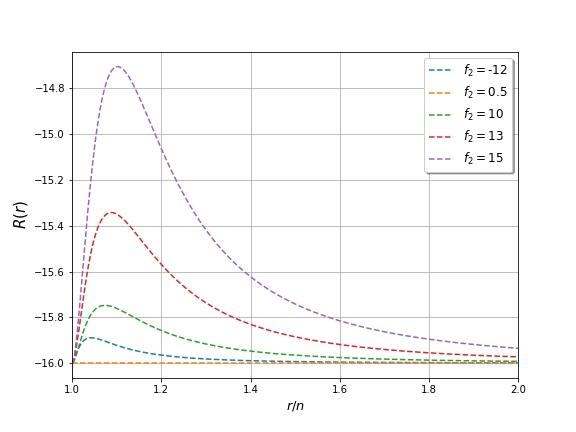}
    \label{Ricciscalarnut}
\end{subfigure}~
\begin{subfigure}[b]{0.5\textwidth}
    \centering
    \includegraphics[scale=0.45]{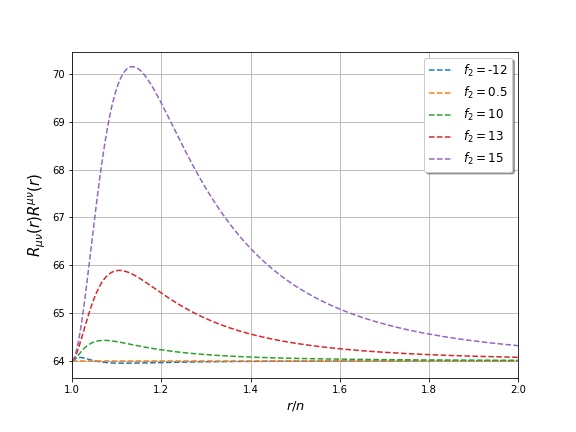}
    \label{Ricci2nut}
\end{subfigure}\\
\begin{subfigure}[b]{0.5\textwidth}
    \centering
    \includegraphics[scale=0.45]{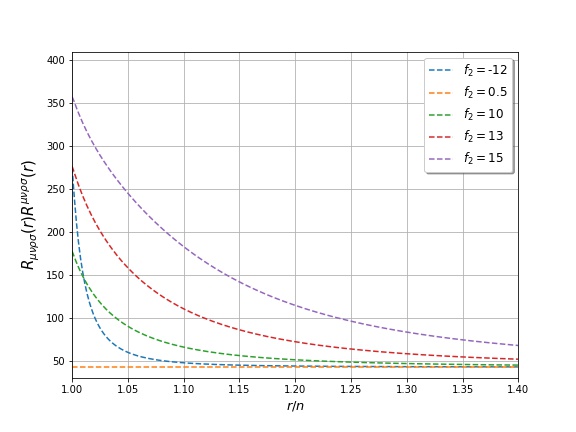}
    \label{Riemann2NUT}
\end{subfigure}
\caption{Curvature invariants for Taub-NUT with $\alpha=4$, $\beta=1$, $\Lambda=-10$, and $n=1$, using different values of $f_2$. The case when $f_2=0.5$ represents the ground state of Eq.~\eqref{fmaxsym} and all the curvature invariants remain constant.}
\label{invariantsnut}
\end{figure}

Then, we proceed to solve Eq.~\eqref{mastereq} numerically. First, we notice that the master equation for $f(r)$ is singular when $r=n$. Therefore, we introduce a regulator $\epsilon\ll1$, such that the initial conditions read
\begin{subequations}\label{icnut}
\begin{align}
f(r=n+\epsilon)&= f_1\left(\epsilon + f_2 \epsilon^2  + f_3 \epsilon^3 + \ldots\right), \\ f'(r=n+\epsilon)&=f_1\left(1+2\epsilon f_2 + 3\epsilon^2 f_3 + \ldots\right).
\end{align}
\end{subequations}
We consider up to cubic order in $\epsilon$ to enhance numerical precision. The metric functions and curvature invariants are plotted in Fig.~\ref{barfnut} and Fig.~\ref{invariantsnut} for different values of $f_2$, respectively.

\begin{figure}
    \centering
    \includegraphics[scale=0.45]{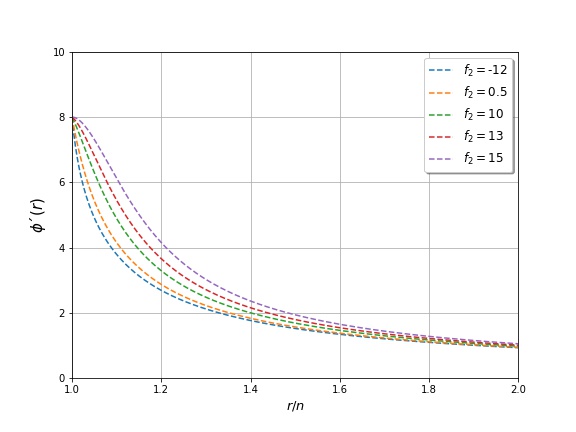}
    \caption{Numerical solution of $\phi'(r)$ [c.f. Eq.~\eqref{phisolnut}] according to the evolution of $f(r)$ in Fig.~\ref{barfnut}. Recall, the energy-momentum tensor of the scalar field gravitates as an effective cosmological constant as $r\to\infty$ since the system approaches asymptotically locally to the Euclidean AdS space in Eq.~\eqref{maxsym}.}
    \label{dphinut}
\end{figure}

For bolt, on the other hand, the metric function admits a series expansion near the horizon $r=r_b$, with $r_b>n$, as
\begin{align}\label{boltexpansion}
    f(r) = \bar{f}_1\left[\left(r-r_b \right) + \bar{f}_2\left(r-r_b \right)^2 + \bar{f}_3\left(r-r_b\right)^3 + \bar{f}_4\left(r-r_b \right)^4 + \ldots \right].
\end{align}
The absence of conical singularities expressed in terms of the regularity conditions~\eqref{forbolt}, relates the free parameter $\bar{f}_1$ with the period of the Euclidean time $\beta_\tau$ through
\begin{align}\label{barf1}
    \bar{f}_1 &= \frac{16\pi^2\beta r_b}{\beta_\tau^2\left[\alpha\left(r_b^2 -n^2 \right) + \beta \right]},
\end{align}
where the solution of $h(r)$ in Eq.~\eqref{hsolnut} has been used. Moreover, unobservability of Misner strings impose the relation $\beta_\tau=8\pi n$ for Taub-Bolt, relating the parameter $\bar{f}_1$ with $r_b$. Thus, the Hawking temperature of this solution is $T= (8\pi n)^{-1}$. Additionally, since $\alpha$ and $\beta$ must be positive for avoidance of instabilities~\cite{Jiang:2017imk} and for consistency with solar system tests~\cite{Gonzalez:2020vzl}, we found that $\bar{f}_1>0$ by virtue of $r_b > n$.

\begin{figure}[t]
\begin{subfigure}[b]{0.5\textwidth}
    \centering
    \includegraphics[scale=0.45]{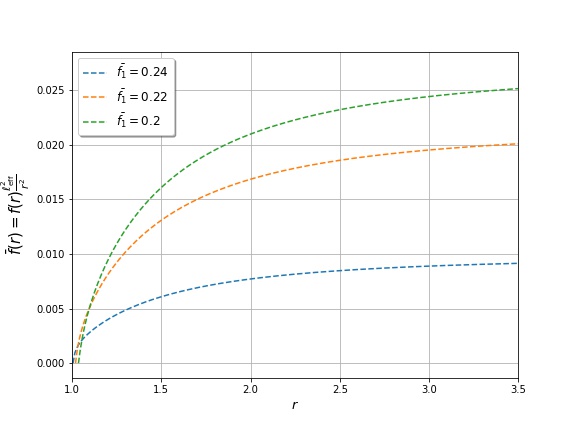}
\end{subfigure}~
\begin{subfigure}[b]{0.5\textwidth}
    \centering
    \includegraphics[scale=0.45]{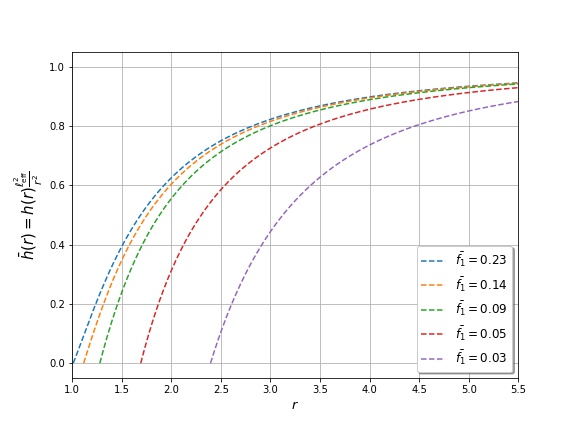}
\end{subfigure}
\caption{Numerical solution of Eq.~\eqref{mastereq} with initial conditions~\eqref{icbolt}. Here, we have used $\bar{f}(r)=f(r)\tfrac{\ell_{\rm eff}^2}{r^2}$, $\bar{h}(r)=h(r)\tfrac{\ell^2_{\rm eff}}{r^2}$, and $\bar{f}_1$ has been defined in Eq.~\eqref{barf1}. The parameters of the theory have chosen as $\alpha=4$, $\beta=1$, $\Lambda=-10$, $n=1$, and therefore, $0<\bar{f}_1<1/4$. Since $r_b>n$, the upper bound of $\bar{f}_1$ is translated into $r_b\to n$ and the lower one to $r_b\to\infty$. The distinct starting points of $\bar{h}(r)$ are related to the fact that the Taub-bolt radius $r_b\to\infty$ as $\bar{f}_1\to0$, and $r>r_b$.}
    \label{barfhbolt}
\end{figure}

\begin{figure}[t]
\begin{subfigure}[b]{0.5\textwidth}
    \centering
    \includegraphics[scale=0.45]{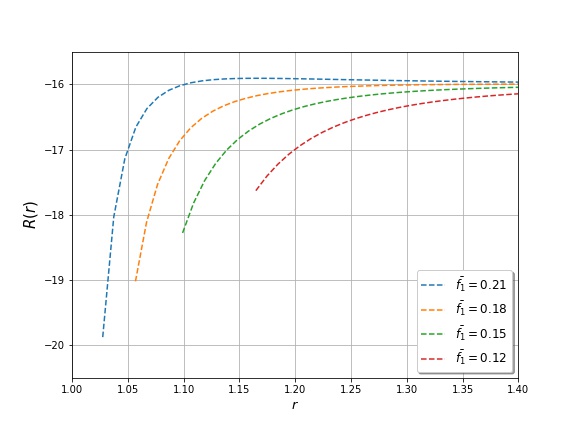}
    \label{Ricciscalarbolt}
\end{subfigure}~
\begin{subfigure}[b]{0.5\textwidth}
    \centering
    \includegraphics[scale=0.45]{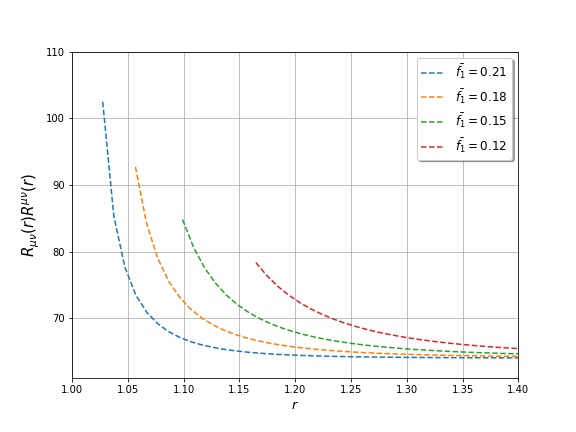}
    \label{Ricci2bolt}
\end{subfigure}\\
\begin{subfigure}[b]{0.5\textwidth}
    \centering
    \includegraphics[scale=0.45]{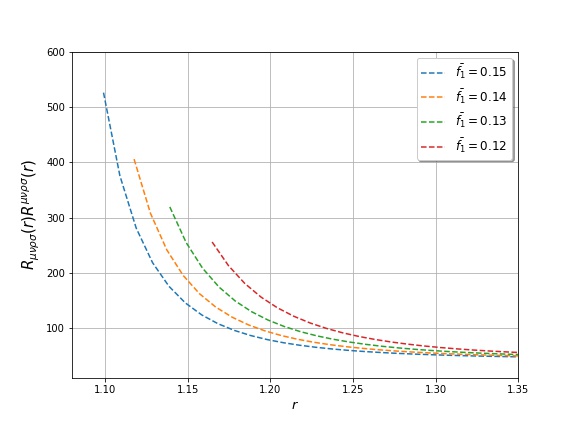}
    \label{Riemann2bolt}
\end{subfigure}
\caption{Curvature invariants for Taub-bolt with $\alpha=4$, $\beta=1$, $\Lambda=-10$, and $n=1$, using different values of $\bar{f}_1$. As explained in Fig.~\ref{barfhbolt}, the different starting points are due to the fact that $r_b\to\infty$ as $\bar{f}_1\to0$ and $r>r_b$. The asymptotically locally AdS behaviour of this solution becomes evident from these plots.}
\label{invariantsbolt}
\end{figure}

Analogous to the NUT case, inserting the series expansion~\eqref{boltexpansion} into Eq.~\eqref{mastereq} and solving for each order of $(r-r_b)$, we find that there is no additional relation between $r_b$ and $n$, in contrast to GR and theories with higher-curvature terms~\cite{Dehghani:2005zm,Dehghani:2006aa,Hendi:2008wq,Bueno:2018uoy,Corral:2019leh}. Moreover, every coefficient $\bar{f}_i$ with $i\geq2$ is obtained recursively in terms of $\bar{f}_1$, and therefore, completely determined in terms of $r_b$ and the parameters of the theory. The leading term is
\begin{align}\notag
\bar{f}_2 &= \frac{1}{2r_b\left(r_b^2-n^2\right)\left[\left(r_b^2-n^2\right)\alpha+\beta\right]\left[\left(r_b^2-n^2\right)\left(\Lambda\beta-\alpha\right)-2\beta\right]}\Big[\alpha^2n^2\left(r_b^2-n^2\right)^2 \\
\notag
&\quad + \alpha\beta\left(r_b^2-n^2\right)\left[n^4\Lambda+6r_b^2+\left(3-\Lambda r_b^2-3r_b\bar{f}_1\right)n^2\right]\\
&\quad +\beta^2\left[\left(r_b \bar{f}_1+1\right)\Lambda n^4-\left(\Lambda r_b^3 \bar{f}_1+3\Lambda r_b^2+2 r_b \bar{f}_1-2\right)n^2+2r_b^4\Lambda+4r_b^2\right]\Big],
\end{align}
and we do not include $\bar{f}_i$ with $i\geq3$ for the sake of simplicity, since they are cumbersome and not very illuminating. The key point is that they are all determined by $r_b$ and the parameters of the theory through $\bar{f}_1$ [cf. Eq.~\eqref{barf1}]. Then, for bolt we solve numerically Eq.~\eqref{mastereq} with initial conditions
\begin{align}\label{icbolt}
    f(r=r_b)=0 \;\;\;\;\; \mbox{and} \;\;\;\;\; f'(r=r_b)= \bar{f}_1,
\end{align}
for different values of $\bar{f}_1$. The numerical evolution of the metric functions and curvature invariants are given in Fig.~\ref{barfhbolt} and Fig.~\ref{invariantsbolt} for different values of $\bar{f}_1$, respectively.

\begin{figure}
    \centering
    \includegraphics[scale=0.45]{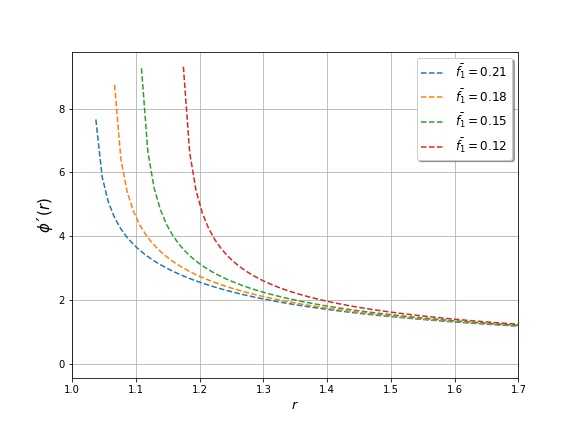}
    \caption{Numerical solution of $\phi'(r)$ [c.f. Eq.~\eqref{phisolnut}] for the bolt case. As in Taub-NUT, the energy-momentum tensor of the scalar field behaves as an effective cosmological constant at infinity since the system approaches asymptotically locally to the Euclidean AdS space in Eq.~\eqref{maxsym}. Additionally, the distinct starting points in this figure are related to the fact that the Taub-bolt radius $r_b\to\infty$ as $\bar{f}_1\to0$, and $r>r_b$, similar to the metric function $\bar{h}(r)$ [cf. Fig~\ref{barfhbolt}].}
    \label{dphibolt}
\end{figure}



To obtain the mass of the solution, we need to know the asymptotic behaviour of the metric function $f(r)$. To this end, we perform the series expansion 
\begin{align}\label{fasymnut}
    f(r) = \frac{r^2}{\ell_\infty^2} + \mu_\infty - \frac{\mu_1}{r} +  \frac{\mu_2}{r^2} + \frac{\mu_3}{r^3} + \ldots,
\end{align}
as $r\to\infty$. Replacing this expression into Eq.~\eqref{hsolnut}, we find that $h(r) = r^2/\ell_{\rm eff}^2 + ...$ at large $r$, independent of the value of $\ell_{\infty}^2$ (recall $\ell_{\rm eff}^{-2} = \tfrac{\alpha}{3\beta}$). Indeed, it yields asymptotically to the Riemann curvature of Eq.~\eqref{Riemannmaxsym} for arbitrary coefficients in the expansion~\eqref{fmaxsym}. Therefore, we conclude that this solution is asymptotically locally AdS with curvature radius $\ell_{\rm eff}^{-2} = \tfrac{\alpha}{3\beta}$ and it generalizes the black hole of Ref.~\cite{Anabalon:2013oea} by introducing the NUT charge. 

Then, we proceed to determine the coefficients of Eq.~\eqref{fasymnut} from the field equations. To do so, we insert the expansion~\eqref{fasymnut} into the master equation~\eqref{mastereq} and solve for each order at large $r$. We find that $\ell_{\infty}$ and $\mu_1$ remain arbitrary, while the other coefficients are fixed in terms of the latter as
\begin{align}
  \mu_\infty &= -\frac{1}{\alpha\ell_\infty^2\left(\Lambda\beta - \alpha \right)}\left[9\alpha\beta + 3\Lambda\beta^2 +\alpha n^2\left(\Lambda\beta-\alpha \right) - \frac{12\beta n^2\left(\Lambda\beta+3\alpha \right)}{\ell_\infty^2} \right], \\
  \mu_2 &= -\frac{3\beta^2}{\left[\alpha\ell_\infty\left(\Lambda\beta-\alpha \right)\right]^2} \left[\left(\Lambda\beta+\alpha \right)^2 - \frac{n^2\left(11\Lambda^2\beta^2 + 46\Lambda\alpha\beta + 59\alpha^2\right)}{\ell_\infty^2} + \frac{4n^4\left(7\Lambda^2\beta^2 + 38\Lambda\alpha\beta + 55\alpha^2 \right)}{\ell_\infty^4} \right],  \\
  \mu_3 &=  -\frac{\mu_1 n^2}{2}\left[1 - \frac{3\beta\left(\Lambda\beta + 3\alpha \right)}{\alpha\ell_\infty^2\left(\Lambda\beta-\alpha \right)} \right],
\end{align}
and so on, where $\Lambda\beta-\alpha\neq0$. If the critical case, i.e. $\Lambda\beta-\alpha=0$, would have been taken beforehand, the system degenerates as shown in Ref.~\cite{Bravo-Gaete:2014haa}. In contrast to the solution of Ref.~\cite{Anabalon:2013oea}, here the presence of the NUT avoids one to eliminate any of these free coefficients through a redefinition of the Euclidean time as it can be seen from the off-diagonal pieces of Eq.~\eqref{euclideanmetricansatz}. However, compatibility of the scalar field with asymptotically AdS symmetry imposes an additional relation between the parameters of the solution~\cite{Hertog:2004dr,Henneaux:2006hk}. Since Taub-NUT/Bolt-AdS determines its own asymptotic behaviour, we conjecture that a similar relation should arise by demanding boundary conditions for the scalar field compatible with the asymptotic symmetries of Eq.~\eqref{euclideanmetricansatz}.  

Next, we employ the formalism of Balasubramanian--Krauss~\cite{Balasubramanian:1999re} to compute the mass. The renormalized boundary stress-energy tensor of Horndeski gravity was obtained in Ref.~\cite{Liu:2017kml} for asymptotically AdS planar black holes coupled to scalar fields possessing radial dependence. For the solution presented here, the base manifold has spherical topology and the renormalized boundary stress-energy tensor is
\begin{align}
    T^{\mu\nu} &= 2\left(\kappa + \frac{\beta}{4}\nabla_\lambda\phi \nabla^\lambda\phi \right)\left(\mathcal{K}h^{\mu\nu} - \mathcal{K}^{\mu\nu} \right) - 2\zeta_2\left(\mathcal{R}^{\mu\nu} - \frac{1}{2}h^{\mu\nu}\mathcal{R} \right) + \zeta_1 h^{\mu\nu},
\end{align}
where the values of $\zeta_{1,2}$ are given in Eq.~\eqref{ctmaxsym} and $\phi=\phi(r)$. This expression is equivalent to the electric part of the Weyl tensor in the pure gravity case~\cite{Ashtekar:1999jx,Miskovic:2009bm} for asymptotically AdS spacetimes. Moreover, it is covariant provided that the intrinsic curvature tensor is expressed through contractions of the Gauss--Codazzi equation as
\begin{align}
    \mathcal{R}_{\mu\nu} &= R_{\mu\nu} - 2n^\lambda R_{\lambda(\mu}n_{\nu)} + R_{\lambda\rho}n^\lambda n^\rho n_\mu n_\nu - R_{\lambda\mu\rho\nu}n^\lambda n^\rho - \mathcal{K}_{\mu\lambda}\mathcal{K}^{\lambda}{}_\nu + \mathcal{K}\mathcal{K}_{\mu\nu}.
\end{align}
We define the unit normal vector $u^\mu$ that generates the flow of Euclidean time in $\partial\mathcal{V}$. The mass is obtained by integrating $T_{\mu\nu}u^\mu \xi^\nu$ over the codimension-2 boundary at infinity, $\Sigma$, where $\xi^\mu$ is the Killing vector associated to Euclidean time symmetry. This yields to\footnote{The minus sign on the right-hand side of the first equality stem from the fact that we are working in Euclidean signature.}
\begin{align}\label{Mnumeric}
    M = -\int_\Sigma \diff{^2x}\sqrt{\sigma}\;  T_{\mu\nu} u^\mu \xi^\nu = -\frac{\mu_1\ell_\infty\left(\Lambda\beta-\alpha \right)}{4G\sqrt{3\alpha\beta}} ,
\end{align}
where $\sigma$ is the determinant of the induced metric on $\Sigma$ (for details see~\cite{Balasubramanian:1999re}). 

Some comments are in order. First, the mass of the solution vanishes in the degenerated sector of the theory, i.e. $\alpha=\Lambda\beta$ (see Ref.~\cite{Bravo-Gaete:2014haa}). Second, positivity of $M$ when $\mu_1\ell_\infty>0$, alongside the absence of ghosts ($\alpha>0$) is guaranteed by reality of scalars fields in the asymptotic region as long as $\Lambda<0$ [see Eq.~\eqref{maxsym}]. Third, although the series expansion of Taub-NUT and Taub-Bolt solutions are different near the fixed points, they have the same asymptotic behaviour as $r\to\infty$. The unobservability of Misner strings, on the other hand, implies that the Hawking temperature in both cases is $T=\left(8\pi n \right)^{-1}$. Remarkably, there is no relation whatsoever between the bolt radius and the NUT charge, in contrast to GR and higher-curvature theories. This stem from the fact that the metric functions $f(r)$ and $h(r)$ are different for nontrivial scalar fields. Thus, the solutions presented here have more free parameters than in GR and higher-curvature gravity and we conclude that this is evidence for scalar hair. 

It is worth mentioning that one can match $M$ with the coefficients $f_2$ and $\bar{f}_1$ by fitting the numerical solutions with appropriate polynomials to read off the coefficients in the asymptotic expansion~\eqref{fasymnut}. For Bolt, we find that $M$ satisfies a quadratic relation in terms of $r_b$. For NUT, however, we find that there exists a range in the parameter $f_2$ where $M$ turns out to be negative. This can be avoided by restricting $1/2<f_2\lesssim4.1$ for the choice of parameters assumed in the numerical integration (for instance see Fig.~\ref{barfnut}).

In the following, we explore higher-dimensional $p$-branes and solitons with NUT charge in Horndeski gravity supported by axionic fields.

\section{Higher-dimensional Taub-NUT/Bolt branes and solitons with axionic fields\label{sec:branes}}

Axionic scalar fields have recently attracted a lot of interest, since they can be used to construct homogeneous black strings in AdS~\cite{Cisterna:2017qrb}. These fields are characterized by nontrivial dependence on the coordinates of the target space and it has been shown that they might avoid the Gregory--Laflamme instability at linear level~\cite{Gregory:1993vy}, regardless the size of the Schwarzschild-AdS black hole located on the brane~\cite{Cisterna:2019scr}. Axions support the existence of homogeneous AdS black strings in Einstein--Gauss--Bonet gravity~\cite{Cisterna:2018mww} and it was shown that they must possess particular kinetic coupling in order for these solutions to exists. Rotating black strings in Chern--Simons modified gravity~\cite{Alexander:2009tp} are supported by axionic fields as well, representing the first analytic rotating solution with contribution of the Cotton tensor~\cite{Cisterna:2018jsx} in such a framework. Here, we extend these solutions by including the NUT charge in the context of Horndeski theory.

In order to construct $D=4+p$ Taub-NUT/Bolt-AdS branes analyticall, we consider the action~\eqref{action} nonminimally coupled to $p$ scalar fields $\psi_i$ with $i=1,\ldots,p$, whose target space is $\mathbb{R}^p$, namely
\begin{align}\label{actionaxion}
    \tilde{I}_{\rm H}\left[g_{\mu\nu},\psi_i\right]&=\int_{\mathcal{V}} d^{4+p} x \sqrt{-g}\left[\kappa(R-2 \Lambda)-\frac{1}{2}\left(\alpha g^{\mu \nu}-\beta G^{\mu \nu}\right)\delta_{ij} \nabla_{\mu} \psi^i \nabla_{\nu} \psi^j\right],
\end{align}
where $\delta_{ij}$ is the $p$-dimensional Kronecker delta. The field equations for this system are similar to Eqs.~\eqref{eom}, but with additional scalar fields, i.e.,
\begin{subequations}\label{eomaxion}
\begin{align}\label{eomgaxion}
\mathcal{E}_{\mu\nu} &\equiv G_{\mu\nu} + \Lambda g_{\mu\nu} - \frac{\alpha}{2\kappa}\mathcal{T}_{\mu\nu}^{(1)} - \frac{\beta}{2\kappa} \mathcal{T}_{\mu\nu}^{(2)} = 0, \\
 \label{eompaxion}
\mathcal{E}_i &\equiv \nabla_\mu \mathcal{J}^\mu_i = 0,
\end{align}
\end{subequations}
where we have defined
\begin{align}
\mathcal{T}_{\mu\nu}^{(1)} &= \delta_{ij} \left( \nabla_\mu\psi^i\nabla_\nu\psi^j - \frac{1}{2}g_{\mu\nu}\nabla_\lambda\psi^i\nabla^\lambda\psi^j\right),\\
\notag
\mathcal{T}_{\mu\nu}^{(2)} &= \delta_{ij} \bigg( \frac{1}{2}\nabla_\mu\psi^i\nabla_\nu\psi^j R - 2\nabla_\lambda\psi^i\nabla_{(\mu}\psi^j R^{\lambda}{}_{\nu)} - \nabla^\lambda\psi^i\nabla^\rho\psi^j R_{\mu\lambda\nu\rho} \\
\notag 
&\quad - \left(\nabla_\mu\nabla_\lambda\psi^i\right)\left(\nabla_\nu\nabla^\lambda\psi^j\right)
 + \left(\nabla_\mu\nabla_\nu\psi^i \right)\Box\psi^j + \frac{1}{2}G_{\mu\nu}\nabla_\lambda\psi^i\nabla^\lambda\psi^j \\
&\quad - \frac{1}{2}g_{\mu\nu}\left[\left(\Box\psi^i\right)\left(\Box\psi^j\right) - \left(\nabla_\lambda\nabla_\rho\psi^i\right)\left(\nabla^\lambda\nabla^\rho\psi^j\right) - 2\nabla^\lambda\psi^i\nabla^\rho\psi^j R_{\lambda\rho} \right]\bigg),\\
\mathcal{J}^\mu_i &= \left(\alpha g^{\mu\nu} - \beta G^{\mu\nu}\right)\nabla_\nu\psi_i.
\end{align}

To solve the field equations, we consider the product space between the family of inhomogeneous Euclidean metrics of Eq.~\eqref{euclideanmetricansatz} and $\mathbb{R}^p$, written in local coordinates as
\begin{align}\label{braneansatz}
    \diff{s^2} = f(r)\left(\diff{\tau} + 2n\mathcal{A}_{(k)}\right)^2 + \frac{\diff{r^2}}{h(r)} + (r^2 - n^2)\diff{\Sigma^2_{(k)}} + \delta_{ij}\diff{z^i}\diff{z^j},
 \end{align}
where the K\"ahler potential $\mathcal{A}_{(k)}$ has been defined in Eq.~\eqref{kahlerpotential}.
Although this metric is homogeneous along the coordinates $z^i$, we do not impose translational symmetry on the scalar fields along $\mathbb{R}^p$, i.e. $\psi_i=\psi_i(r,z^i)$. The Klein--Gordon equations~\eqref{eompaxion} are solved by scalar fields with linear dependence on the coordinates that span the brane, that is
\begin{align}\label{axionsolbrane}
    \psi^i &= \lambda z^i,
\end{align}
where $\lambda$ is an integration constant usually referred to as the axionic charge. Remarkably, the on-shell energy-momentum tensors associated to the solution~\eqref{axionsolbrane} has the same isometries of the line element~\eqref{braneansatz}, despite the linear dependence of the scalar fields on the local coordinates of $\mathbb{R}^p$. This is a consequence of the shift symmetry in the scalar fields' space and it can be seen as a Freund--Rubin compactification~\cite{Freund:1980xh} where the internal manifold is flat and, instead of magnetic gauge fields, scalar fields support their existence.

The ansatz~\eqref{braneansatz} alongside the scalar profile~\eqref{axionsolbrane} allows one to decompose the field equation for the metric as
\begin{align}\label{transverseeom}
    \mathcal{E}_{\bar{\mu}\bar{\nu}} &= \left[1 - \frac{p\beta\lambda^2}{4\kappa} \right]G_{\bar{\mu}\bar{\nu}} + \left[\Lambda + \frac{p\alpha\lambda^2}{4\kappa} \right]g_{\bar{\mu}\bar{\nu}} = 0, \\
    \mathcal{E}_{ij} &= -\frac{1}{2}\left[1-\frac{(p-2)\beta\lambda^2}{4\kappa} \right]\delta_{ij}R + \left[\Lambda + \frac{(p-2)\alpha\lambda^2}{4\kappa} \right]\delta_{ij} = 0,
\end{align}
where barred greek and latin characters denote components of transverse and brane sections, respectively. Taking the trace on both equations, i.e. $g^{\bar{\mu}\bar{\mu}}\mathcal{E}_{\bar{\mu}\bar{\nu}}=0$ and $\delta^{ij}\mathcal{E}_{ij} = 0$, one obtains
\begin{subequations}\label{Rcompatibility}
\begin{align}
    R &= \frac{4\left[4\Lambda\kappa + p \alpha \lambda^2\right]}{4\kappa - p\beta \lambda^2}, \\
    R &= \frac{2\left[4\Lambda\kappa + (p-2) \alpha\lambda^2 \right]}{4\kappa - (p-2)\beta\lambda^2}.
\end{align}
\end{subequations}
Compatibility of these two equations gives rise to a quartic equation for the axionic charge whose solution, for $p\neq2$, fixes the latter in terms of the parameters of the theory as\footnote{The case when $p=2$ fixes axionic charge according to $\lambda^2=-\tfrac{2\kappa\Lambda}{\Lambda\beta+2\alpha}$.}
\begin{align}    \label{lambda2sol}
    \lambda^2_\pm &= \frac{2\kappa}{\alpha\beta p(p-2)}\bigg[\alpha(p+2) - \Lambda\beta(p-4) \pm \sqrt{\beta^2(p-4)^2\Lambda^2+2\alpha\beta\Lambda \left(p^2-2p+8\right)+\alpha^2(p+2)^2} \bigg].
\end{align}
Reality of the scalar fields impose conditions on the parameters of the theory since $\lambda_\pm^2>0$. Moreover, the argument of the square root must be either positive or zero. When positive, these conditions imply that $\alpha>0$, $\Lambda<0$, and that $\lambda_{\pm}$ must lie in the range provided in Table~\ref{Conditions for lambda^2>0} for different values of $p$. Here, for $p\neq4$, the parameter $\rho_\pm$ has been defined as
\begin{equation}\label{rhopmdef}
  - \frac{\Lambda\rho_{\pm}}{\alpha} = \frac{\left(p^2-2p+8\right) \pm  \sqrt{32(p-2) p }}{(p-4)^{2}}.
\end{equation}
\begin{table}
    \centering
    \begin{tabular}{|c|c|c|c|}
        \hline
        $0<p<2$   & $\beta>0$   &   $\lambda_-^2>0$ &  -\\
         $2<p<4$    & $0<\beta<\rho_-$   & $\lambda_-^2>0$ & $\lambda_+^2>0$ \\
           $p=4$     &    $0<\beta<- \frac{9 \alpha}{8 \Lambda}$     & $\lambda_-^2>0$ & $\lambda_+^2>0$ \\
           $p>4$    & $0<\beta<\rho_- \cup \beta> \rho_+$ & $\lambda_-^2>0$ & $\lambda_+^2>0$\\
           \hline
    \end{tabular}
    \caption{Conditions on the parameters of Horndeski theory for the existence of Taub-NUT/Bolt-AdS $p$-branes in different dimensions. Notice that the first case is allowed only when $\lambda_-^2>0$. The definition of $\rho_\pm$ is given in Eq.~\eqref{rhopmdef}.}
    \label{Conditions for lambda^2>0}
\end{table}

When $4\kappa-p\beta\lambda_\pm^2\neq0$, the transverse components of the field equation for the metric can be cast into the form
\begin{align}\label{eomgtrans}
    \mathcal{E}_{\bar{\mu}\bar{\nu}} = G_{\bar{\mu}\bar{\nu}} + \Lambda_{\rm eff}\,g_{\bar{\mu}\bar{\nu}} = 0, 
\end{align}
which resembles Einstein--AdS equations with an effective cosmological and gravitational constants given by 
\begin{align}\label{Lambdaeff}
    \Lambda_{\rm eff} = \frac{4\kappa\Lambda + p\alpha\lambda_\pm^2}{4\kappa_{\rm eff}} \;\;\;\;\; \mbox{and} \;\;\;\;\; \kappa_{\rm eff}=\kappa - \frac{p\beta\lambda_\pm^2}{4},
\end{align}
respectively. The appearance of an effective Newton constant is a consequence of the nonminimal coupling of the scalar fields to the Einstein's tensor in Horndeski theory. Solving Eq.~\eqref{eomgtrans}, one obtains that the metric functions are
\begin{align}\label{fsolbrane}
    f(r) &= h(r) =  k\left(\frac{r^2 + n^2}{r^2-n^2}\right) -\frac{2MG_{\rm eff} r}{r^2-n^2} - \frac{\Lambda_{\rm eff}}{3}\frac{\left(r^4 - 6n^2 r^2 - 3n^4 \right)}{r^2-n^2},
\end{align}
where $M$ is an integration constant and $G_{\rm eff}$ can be read from Eq.~\eqref{Lambdaeff} as 
\begin{align}
G_{\rm eff} &= \frac{G}{1-4\pi Gp\beta\lambda_\pm^2}.
\end{align}
Regularity conditions~\eqref{boundaryconditions} on the hypersurfaces at constant $z^i$, implies that the integration constant $M$ is fixed according to Eqs.~\eqref{MnutGR} or~\eqref{MboltGR}, for nuts and bolts, respectively, with the replacement $(G,\Lambda)\to(G_{\rm eff},\Lambda_{\rm eff})$. Moreover, when $k=1$, the unobservability of Misner strings on the transverse section implies that the period of Euclidean time is fixed as $\beta_\tau = 8\pi n$. This, in turn, impose a relation between the radius of the Killing horizon and the NUT charge according to Eq.~\eqref{rbsol} with an effective cosmological constant dressed by scalar fields as~\eqref{Lambdaeff}.

To compute the thermodynamic properties of the solution with metric functions given in Eq.~\eqref{fsolbrane}, we focus on spherically symmetric transverse section, i.e. $k=1$. First, we notice that the scalar fields with linear profile generate an energy-momentum tensor that behaves as a cosmological constant, producing an asymptotically locally Euclidean AdS$_4\times\mathbb{R}^p$ space. Thus, the renormalized Euclidean action is obtained when the counterterms~\eqref{ICT} are chosen such that 
\begin{align}\label{zeta1sol}
\zeta_1 &= -\frac{1}{2}\sqrt{-\frac{\Lambda_{\rm eff}}{3}}\bigg(8\kappa + p\beta\lambda_\pm^2 \bigg)  ,\\
\label{zeta2sol}
    \zeta_2 &= -\frac{1}{8}\sqrt{-\frac{3}{\Lambda_{\rm eff}}}\bigg(8\kappa-p\beta\lambda_\pm^2 \bigg) ,
\end{align}
respectively. Of course, this does not mean that the counterterms depend on the integration constants of the solution since, recall, $\lambda_\pm^2$ is fixed on shell in terms of the parameters of the theory and the dimensionality of spacetime through Eq.~\eqref{lambda2sol}. Thus, the renormalized Euclidean action for the Taub-NUT/Bolt-AdS $p$-brane with axionic profile in Horndeski gravity is given by 
\begin{align}\label{Irenvalue}
 I_{\rm ren} &= \frac{8\pi \beta_\tau \kappa_{\rm eff}V}{3}\left(3MG_{\rm eff} - 3n^2 r_b \Lambda_{\rm eff} + r_b^3\Lambda_{\rm eff} \right),
\end{align}
where $V$ is the volume of $\mathbb{R}^p$. For bolt, we obtain that the free energy, mass, entropy, and specific heat are
\begin{align}
    \mathcal{F}_{\rm bolt} &= \frac{4\pi \kappa_{\rm eff}V}{r_b} \left[ r_b^2 + n^2 + \frac{\Lambda_{\rm eff}}{3}\left(r_b^4 + 3n^4 \right) \right], \\
    \mathcal{M}_{\rm bolt} &= \frac{8\pi\kappa_{\rm eff}V}{r_b}\left[\left(r_b^2+n^2\right)\left(1+n^2\Lambda_{\rm eff}\right)-\frac{\Lambda_{\rm eff}r_b^2}{3}\left(r_b^2-3n^2\right) \right],\\
    \notag
    \mathcal{S}_{\rm bolt} &=  \frac{32\pi^2 n\kappa_{\rm eff}V}{r_b^2\left(1+4nr_b\Lambda_{\rm eff} \right)}\bigg[r_b\left(r_b^2+n^2 \right) + \Lambda_{\rm eff}\left(r_b^5 + 12r_b^2 n^3 + 3 r_b n^4 - 4n^5 \right) \\
    &\qquad  + 4\Lambda_{\rm eff}^2n^3\left(r_b^4+4r_b^2 n^2-n^4 \right)  \bigg],\\
    \notag
    \mathcal{C}_{\rm bolt} &= -\frac{64\pi^2 n \kappa_{\rm eff}V}{r_b^3\left(1+4 n r_b\Lambda_{\rm eff} \right)^3}\bigg[r_b^2\left(r_b^2+n^2\right) +2r_b \Lambda_{\rm eff}\left(r_b^5+ r_b^4 n + 16 r_b^2 n^3 + 3r_b n^4 - 5n^5 \right)\\
    \notag
  & \qquad + 2n\Lambda_{\rm eff}^2\left(r_b+n \right)\left(3 r_b^6- 3 r_b^5 n + 12 r_b^4 n^2 + 68r_b^3 n^3-23r_b^2 n^4-17r_b n^5 + 8n^6 \right)\\
  \notag
  &\qquad + 16n^4\Lambda_{\rm eff}^3\left(5r_b^6 + 18 r_b^5 n + 27 r_b^4 n^2 - 16 r_b^3 n^3 - 9r_b^2 n^4+6 r_b n^5 + n^6  \right)\\
  &\qquad+32 r_b n^5 \Lambda_{\rm eff}^4 \left(3 r_b^6 +21 r_b^4 n^2 - 11 r_b^2 n^4 + 3n^6 \right)\bigg],
\end{align}
respectively. The NUT case can be obtained by taking the limit $r_b\to n$ on these expressions, giving
\begin{align}
    \mathcal{F}_{\rm nut} &= 8\pi n\kappa_{\rm eff}V\left(1 + \frac{2n^2 \Lambda_{\rm eff}}{3} \right), \\
    \mathcal{M}_{\rm nut}  &=  16\pi n\kappa_{\rm eff}V\left( 1 + \frac{4}{3}\Lambda_{\rm eff}n^2\right), \\
    \mathcal{S}_{\rm nut} &=  64\pi^2 n^2 \kappa_{\rm eff}V\left(1+2n^2\Lambda_{\rm eff} \right),\\
    \mathcal{C}_{\rm nut} &= -128\pi^2 n^2\kappa_{\rm eff} V\left(1 + 4n^2\Lambda_{\rm eff} \right),
\end{align}
where the contribution of the Misner string in both cases becomes evident. It is straightforward to check that the mass density reproduces the result of Eqs.~\eqref{MnutGR} and~\eqref{MboltGR} for nuts and bolts, respectively, by performing the substitution $(G,\Lambda)\to(G_{\rm eff},\Lambda_{\rm eff})$. Moreover, we corroborated that these expressions satisfy the first law of thermodynamics, namely
\begin{align}
    \diff{\mathcal{M}} = T\diff{\mathcal{S}}.
\end{align}
Notice that the axionic charge does not enter here since its value is fixed on shell in terms of the parameters of the theory. In order to have a positive mass, entropy, and specific heat, the nut charge must satisfy
\begin{align}
    \sqrt{-\frac{1}{4\Lambda_{\rm eff}}}<n<\sqrt{-\frac{1}{2\Lambda_{\rm eff}}}.
\end{align}

Finally, we notice that a regular soliton can be obtained for $p=1$, by performing a Wick rotation $z\to -it$ on the metric~\eqref{braneansatz}, giving
\begin{align}\label{soliton}
    \diff{s^2} = -\diff{t^2} + f(r)\left(\diff{\chi} + 2n\mathcal{A}_{(k)}\right)^2 + \frac{\diff{r^2}}{f(r)} + (r^2 - n^2)\diff{\Sigma^2_{(k)}}.
 \end{align}
Here, the metric function $f(r)$ is given by Eq.~\eqref{fsolbrane} and it is supported by a time-dependent scalar field $\psi=-\lambda_\pm it$, whose reality condition implies that $\lambda_\pm^2<0$ [cf. Eq.~\eqref{lambda2sol}]. Remarkably, we found that there exists region on the parameter space where this condition is fulfilled for $\alpha,\beta>0$, namely (i) $\lambda_-^2<0$ for $\Lambda>0$, and (ii) $\lambda_+^2<0$ for $\Lambda\in\mathbb{R}$.  

To compute the mass of the soliton, we first note that there is no horizon associated to the time-like Killing vector $\xi=\partial/\partial t$. Since entropy arises from obstructions to foliating the spacetime with hypersurfaces of constant $t$ due to the presence of fixed points, it is direct to see that the entropy of this regular soliton vanishes. Therefore, the mass can be obtained by multiplying the Euclidean on-shell action by the inverse of an arbitrary period of Euclidean time, as a consequence of the Gibbs--Duhem relation $S=\beta_\tau M - I_{\rm reg}$~\cite{Ghezelbash:2001vs,Clarkson:2003wa}. In fact, it was shown that this method coincides with the conserved charge associated to the boundary stress-tensor of the Eguchi--Hanson soliton in five-dimensional GR~\cite{Clarkson:2005qx,Clarkson:2006zk}. For $k=1$, the mass of the soliton is
\begin{align}
    \mathscr{M}_{\rm nut} &=  64\pi^2 n^2 \kappa_{\rm eff}\left(1+\frac{2}{3}n^2\Lambda_{\rm eff} \right) , \\
    \mathscr{M}_{\rm bolt} &= \frac{32\pi^2 n \kappa_{\rm eff}}{r_b}\left[r_b^2 + n^2 + n^4\Lambda_{\rm eff} + \frac{1}{3}r_b^3 \Lambda_{\rm eff} \right],
\end{align}
for nuts and bolts, respectively. 

Some remarks are now in order. First, we notice that there is no analog in five-dimensional GR with cosmological constant. This stem from the fact that the compatibility of the field equations would demand the vanishing of the latter, as it can be seen from Eq.~\eqref{Rcompatibility}. Thus, the presence of scalar fields is crucial for its existence. Second, this odd-dimensional solution can be interpreted locally as the direct product between $\mathbb{R}$ and the $U(1)$ fibration over a two-dimensional K\"ahler manifold. Moreover, this configuration cannot be obtained from the analytical continuation of the solutions presented in Ref.~\cite{Mann:2003zh} due to the absence of a warped factor in the product space and, therefore, it represents a new everywhere regular solution supported by Horndeski scalars. Finally, we notice that when $\Lambda_{\rm eff}<0$ and $n>\sqrt{-3/(2\Lambda_{\rm eff})}$, the mass of the NUT case is negative and this solution has lower energy than Euclidean AdS$_4\times\mathbb{R}$. This result represents an additional example of the one found in Ref.~\cite{Clarkson:2006zk} where a soliton with less energy than global AdS space was obtained.

\section{Conclusions and further remarks\label{sec:conclusions}}

The Taub-NUT/Bolt-AdS solutions bear a close resemblance with instantons in Yang--Mills theory. They represent regular stationary Euclidean configurations whose vacuum is characterized by topologically inequivalent sectors. For the NUT case, the Weyl tensor is globally self dual; the latter being the AdS curvature for Einstein spaces. Their distinct properties have been widely explored and several applications in theoretical physics and differential geometry have emerged. In this work, we show their existence in Horndeski gravity: the most general scalar-tensor theory with second-order field equations.

To do so, we start by solving the Horndeski scalar's zero mode analytically on a Taub-NUT/Bolt background. The energy density in the test-field limit vanishes at the fixed points, however, the norm of the scalar current becomes divergent. To circumvent this problem, we take into account their backreaction and impose regularity. This condition implies that the radial component of the conserved current vanishes and, to avoid the no-hair theorem of Ref.~\cite{Hui:2012qt}, a relation between the metric functions arises. The system reduces to Eq.~\eqref{mastereq}, representing the master equation of the metric function $f(r)$ that we solve for nuts and bolts. 

First, a locally Euclidean AdS space with a nontrivial scalar field is found, whose energy-momentum tensor gravitates as the cosmological constant. To obtain the thermodynamic properties, we compute the Euclidean on-shell action by adding proper counterterms~\cite{Emparan:1999pm} and found that the mass and entropy are zero and constant, respectively. This ground state cannot be obtained continuously from global AdS, due to the presence of a stealth-like scalar field.     

Afterward, we solve the system numerically and find asymptotically locally AdS solutions with NUT charge. Performing series expansion near the fixed points, we obtain precise conditions under which the numerical solutions deviate from the locally AdS with self-gravitating scalar field. The mass is obtained by integrating the renormalized boundary stress-energy tensor over the boundary at infinity [see Eq.~\eqref{Mnumeric}]. Reality of the scalar field and positivity of the mass impose that $\beta>-\alpha/\Lambda$ and $\Lambda<0$. It is worth mentioning that the solutions presented here have more free parameters than in GR, leading us to conclude the presence of scalar hair.

In higher dimensions, we obtain a $p$-brane solution described by the product space between the Hopf-fibered K\"ahler manifold and $\mathbb{R}^p$, which is supported by $p$ Horndeski scalars with axionic profiles. The first law of thermodynamics is satisfied and constraints on the NUT charge appear from the positivity of mass, entropy, and specific heat. Moreover, there exists a particular region in the parameter space that admits a solitonic solution obtained from the analytical continuation of the coordinate that span the brane with $p=1$. This everywhere regular configuration has zero entropy, nontrivial mass, and its $U(1)$ fibration over the K\"ahler manifold belongs to the transverse section. 

Interesting questions remain open. For instance, it is well-known that scalar fields with axionic profile can be used to construct holographic models with momentum relaxation~\cite{Andrade:2013gsa,Jiang:2017imk}. It is worth analyzing condensed matter systems with vorticity from the AdS/CFT viewpoint, since it has been conjectured to be the holographic dual of the NUT charge. Moreover, studying the influence of the latter in the holographic heat current is certainly valuable, since it can be obtained directly from the Noether procedure as shown in Ref.~\cite{Liu:2017kml}. Additionally, the holographic two-point function of dual quantum fields theories can computed by introducing the NUT charge to describe quantum field theories with vorticity, following the prescription of Ref.~\cite{Li:2018rgn}. On the other hand, the relation between the electric part of the Weyl tensor and the boundary stress-energy tensor has been explored for asymptotically AdS spacetimes~\cite{Ashtekar:1999jx,Miskovic:2009bm}. Since the Taub-NUT/Bolt-AdS solution is asymptotically locally AdS and it has magnetic components of the Weyl tensor, their relation to the boundary stress-tensor is worth exploring.  Finally, the relation between holographic renormalization and kounterterms has been recently clarified in Ref.~\cite{Anastasiou:2020zwc}. The role of the latter in regularizing the entropy of the Misner string in spacetimes with AdS asymptotics is relevant and we left this for a forthcoming publication. 

\begin{acknowledgments}
The authors thank to N.~C\'aceres, A.~Cisterna, C.~Erices, D.~Flores-Alfonso, R.~Olea, J.~Oliva, G.~Rubilar, and R.~Stuardo for fruitful discussions, comments, and insightful remarks. E. A. is supported by Universidad de Concepción through Undergraduate-Postgraduate Linkage Scholarship. The work of C. C. is supported by Agencia Nacional de Investigaci\'on y Desarrollo (ANID) through the Fondecyt grant No $11200025$. J. F. acknowledges the financial support of ANID through the Fellowship 22191705. The work of L. S. is supported by 2019/13231-7 FAPESP/ANID.
\end{acknowledgments}

\bibliography{references}

\end{document}